# Behavior of nanoparticle clouds around a magnetized microsphere under magnetic and flow fields

C. Magnet[1], P. Kuzhir[1], G. Bossis[1], A. Meunier[1], S. Nave[1], A. Zubarev[2], C. Lomenech[3] and V. Bashtovoi[4]

[1]University of Nice-Sophia Antipolis, CNRS, Laboratory of Condensed Matter Physics, UMR 7336, 28 avenue Joseph Vallot, 06100, Nice, France

[2] Department of Mathematical Physics, Ural Federal University, 51, Prospekt Lenina, Ekaterinburg 620083 Russia

[3] University of Nice-Sophia Antipolis, Laboratory ECOMERS (Ecosystèmes Côtiers Marins et Réponses aux Stress), EA 4228, 28 avenue Valrose, 06108 Nice Cedex 2, France.

[4] UNESCO Chair "Energy Conservation and Renewable Energies", Belarusian National Technical University, 65, Prospekt Nezavisimosti, 220013 Minsk, Belarus

**Abstract**

When a micron-sized magnetizable particle is introduced into a suspension of nanosized magnetic particles, the nanoparticles accumulate around the microparticle and form thick anisotropic clouds extended in the direction of the applied magnetic field. This phenomenon promotes colloidal stabilization of bimodal magnetic suspensions and allows efficient magnetic separation of nanoparticles used in bioanalysis and water purification. In the present work, size and shape of nanoparticle clouds under the simultaneous action of an external uniform magnetic field and the flow have been studied in details. In experiments, dilute suspension of iron oxide nanoclusters (of a mean diameter of 60 nm) was pushed through a thin slit channel with the nickel microspheres (of a mean diameter of 50μm) attached to the channel wall. The behavior of nanocluster clouds was observed in the steady state using an optical microscope. In the presence of strong enough flow, the size of the clouds monotonically decreases with increasing flow speed in both longitudinal and transverse magnetic fields. This is qualitatively explained by enhancement of hydrodynamic forces washing the nanoclusters away from the clouds. In the longitudinal field, the flow induces asymmetry of the front and the back clouds. To explain the flow and the field effects on the clouds, we have developed a simple model based on the balance of the stresses and particle fluxes on the cloud surface. This model, applied to the case of the magnetic field parallel to the flow, captures reasonably well the flow effect on the size and shape of the cloud and reveals that the only dimensionless parameter governing the cloud size is the ratio of hydrodynamic–to–magnetic forces – the Mason number. At strong magnetic interactions considered in the present work (dipolar coupling parameter $\alpha \geq 2$), the Brownian motion seems not to affect the cloud behavior.

I. Introduction

Colloidal mixture of bimodal charged particles may exhibit a haloing phenomenon characterized by formation of thin clouds of small nanoparticles accumulated around bigger micron-sized particles. This phenomenon is attributed to the interplay between electrostatic and van der Waals interactions between the particles and ensures colloidal stability of the suspension [1-4]. However, in such systems, the cloud thickness is only a few nanometers [5], that allows maintaining a good dispersion state of the suspension only within a narrow range of concentrations of both species.

Much thicker clouds appear in magnetic bimodal suspensions. An external magnetic field magnetizes large micron-sized particles, which attract small superparamagnetic nanoparticles, and the latter form thick anisotropic clouds extended at a distance of a few microparticle diameters in the direction of the applied field. At strong enough magnetic interactions, the ensemble of nanoparticles may undergo a gas-liquid or gas-solid phase transition and condense into highly concentrated domains (clouds) adhered to the microparticle surface [6-7]. Such a phase transition has been proved to enhance significantly the capture efficiency of nanoparticles by magnetic microparticles. On the other hand, the nanoparticle clouds may completely screen dipole-dipole attraction between two micron-sized particles (with dipole moments oriented along the line connecting their centers) and even result in their effective repulsion. This effect has been explained by the interplay between local field modification due to the cloud formation around a pair of microparticles and the osmotic pressure induced by the nanoparticles [8].

Such a field-induced haloing accompanied with a condensation phase transition has at least two potential applications. First, it significantly improves colloidal stability of magnetorheological fluids based on bimodal magnetic particles [9]. Second, in the domain of magnetic filtration, it is expected to broaden the size range of captured particles from micron-sized particles to nanoparticles. This could be an important breakthrough for biotechnology and magnetically assisted water purification [10-12]. Both applications require detailed study of the behavior of nanoparticle clouds around a magnetized microsphere both under flow and in the presence of a magnetic field.

Up to now, theoretical investigations of the magnetic particle capture have been principally motivated by the development of magnetic separation technology. Usually accumulation of magnetic particles around a single magnetized wire or an ordered array of wires was considered. Capture cross-section along with the size and shape of magnetic particle deposits around a magnetized collector were determined. Two distinct approaches were used depending on the size of magnetic particles, or rather on the Péclet number (defined as a ratio of the hydrodynamic – to Brownian forces). For large enough non Brownian particles at large Péclet numbers, the mechanistic approach was employed on the basis of the balance of forces and torques acting on the particles. The capture cross-section was determined via the particle trajectory analysis while the size and the shape of the particle deposits were found from the mechanical equilibrium of the particles on the deposit surface, helpful reviews being given by Gerber and Birss [13], Svoboda [14]. For smaller Brownian particles at low-to-intermediate Péclet numbers, statistical approach was used on the basis of either the convection-diffusion equation or the Langevin equation of particle motion. The former equation gave concentration profiles of the magnetic particles [15-17]. The latter equation was integrated at fixed small time steps to obtain stochastic particle trajectories [18]. Both methods allowed calculation of the capture cross-section as function of the suspension speed and magnetic field strength. However, the steady-state size and shape of the nanoparticle clouds were only found in the limits of flow-dominated (infinite Péclet number) and diffusion-dominated (zero Péclet number) regimes [13, 19-21]. Recently, a quite rigorous approach has been proposed by Chen et al. [22] who have considered the dynamic growth of the nanoparticle clouds as a moving boundary problem, with the field and the flow fields computed numerically. However, this model as well as most existing theories, did not take into account interactions between magnetic nanoparticles that might lead to underestimation of the capture efficiency and even to unphysical results like particle concentrations above the limit of the maximum packing fraction. A few attempts [19, 23, 24] to account for interparticle interactions in the problem of magnetic separation were restricted to non-Brownian particles and did not predict condensation phase transition, which is often observed in magnetic colloids [25-29].

Experimental investigations of the magnetic separation were mostly focused on visualization of the particle trajectories around a magnetized collector, see review by Gerber and Birss [13]. On the other hand, the size and morphology of particle deposits (or clouds) were scarcely studied. Some results were briefly reported for the limits of diffusion-dominated and flow dominated regimes, for which condensation phase transitions were not observed [15, 30, 31]. Furthermore, the studies of particle deposits were restricted to some limited set of experimental parameters and general relationships in terms of dimensionless numbers were not established. Recently, Ivanov and Pshenichnikov [32] have studied a rapid dynamics of accumulation of ferrofluid nanoparticles around a magnetized collector. The authors claim that the nanoparticles undergo the condensation phase transition around a collector and demonstrate a strong recirculation flows induced by the nanoparticle migration towards the collector. However these studies have been carried out in the absence of the external flow, so the flow effect on the behavior of the condensed magnetic phase is still unknown.

In view of the lack of information on this topic and its practical and fundamental interest, we have performed a detailed experimental study of the steady-state behavior of nanoparticle clouds accumulated on the single spherical microspheres in the presence of an external flow and an external magnetic field either aligned or transverse to the flow. To this purpose, we pushed a dilute suspension of magnetic nanoclusters through a microfluidic slit channel, and visualized nanocluster condensation and formation of dense solid-like clouds around a microsphere rigidly attached to one of the channel walls. Experiments have been done in a wide range of suspension velocities. The cloud size and shape have been analyzed as function of the Mason number defined as a ratio of hydrodynamic-to-magnetic forces. For a better understanding of the Mason number effect on the steady-state cloud behavior, we have developed a theoretical model based on the stress balance and particle flux balance on the cloud surface. In the limit of small filtration speeds, phase equilibrium between a solid-like particle cloud and a surrounding medium has been assumed and the nanoparticle concentration inside the clouds has been estimated from the condition of homogeneity of the chemical potential of nanoparticles.

The present article is organized as follows. In the next section II, we present experimental techniques. An overview of visualization results is presented in section III. The size and shape of nanoparticle clouds is analyzed in Section IV in comparison with theoretical estimations. Finally the conclusions and perspectives are outlined in Section V.

**II. Experimental**

The experimental cell used for visualization of nanoparticle clouds around a magnetized microsphere is shown in Fig. 1. A dilute aqueous suspension of iron oxide nanoparticles (at volume fraction equal to 0.32%) was pushed through a slit channel by a syringe pump (KD Scientific KDS 100 Series) at imposed flow rates varying from $Q=7\cdot 10^{-3}$ to 0.14 mL/min. This flow rate corresponds to the flow speeds, $v_0 = Q/S$, varying in the range $1.67\cdot 10^{-4} \leq v_0 \leq 1.79\cdot 10^{-3}$ m/s, with $S$ being the cross-section area of the channel. These speeds correspond to the Reynolds number $Re \ll 1$ at the microsphere scale that implies a laminar flow in this scale.

The flow channel was fabricated by squeezing of a silicon joint (GEB Silicone) between a flat Plexiglas substrate and a microscopic glass plate. Before manufacturing of the channel, spherical nickel microparticles (Alfa Aesar, 300 mesh, 99.8%, sieved to obtain the size ranging from 40 to 50 μm) were attached to the glass plate by heating at 700°C in an oven during two hours. Such a treatment did not cause a significant immersion of the microparticles into the glass plate but ensured a strong enough adhesion so that the particles did not move under suspension flux. The channel dimensions in the direction of the flow (length), fluid vorticity (width) and velocity gradient (height) were 60mm x 10mm x 70±5 μm, respectively. The channel height was measured by an optical microscope and its constancy along the channel walls was approximately adjusted by screws squeezing the glass plate to the Plexiglas substrate through the silicon joint. The flow channel was placed in the transmitted light microscope (Carl Zeiss Photomicroscope III) equipped with a camera PixelInk PL-B742U having a complementary metal oxide semiconductor (CMOS) color image sensor. A 20-fold objective (Olympus IC 20) was used for observations. A stationary magnetic field of an intensity $H_0$=12 kA/m was applied by a pair of Helmholtz coils placed around a microscope and bearing iron yokes, as shown in Fig.1. Measurements showed that the magnetic field was homogeneous within a few percent tolerance in the location of the flow channel. The flow channel was put either along or perpendicularly to the coil axis, so, the magnetic field was either parallel to the flow (longitudinal field) or perpendicular to the flow and parallel to the fluid vorticity (transverse field).

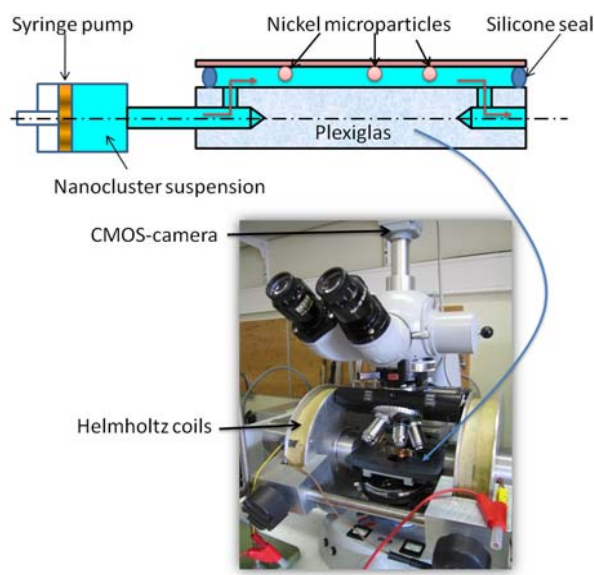

Fig.1. (Color online) Sketch of the experimental setup

The following experimental protocol was adapted. Firstly, the suspension of magnetic nanoparticles was introduced into the syringe pump, which was then connected to the flow channel. The latter was filled with the suspension by pushing the syringe at a speed small enough to avoid entrapment of air bubbles. Once the channel was placed in the microscope, an external magnetic field of a chosen intensity was applied and the system was left at rest for ten minutes. During this time, some magnetic nanoparticles were attracted to nickel microparticles and formed clouds extended along the magnetic field direction. Then, the syringe pump was switched on and the suspension was pushed through the channel at a constant imposed flow rate during two hours. During this time, snapshots of the microparticle with nanoparticle clouds accumulated around were taken with a one minute interval, and three videos of a few-minute duration were recorded in the beginning, in the middle and at the end of the observation process. After the flow onset, we observed a rapid partial destruction of the clouds under hydrodynamic forces followed by their reconstruction on a time scale of about one hour. After this time, a quasi-steady state regime was achieved so that the cloud size, shape and morphology did not evolve significantly. At the end of the observation period, the field was switched off, the flow was stopped, the syringe pump was filled with a new portion of the magnetic suspension, and the experience was repeated at another flow rate. To check the reproducibility, all the measurements were conducted two times for the same set of experimental parameters. The steady-state shape of the clouds (at an elapsed time equal to two hours from the flow onset) was analyzed and quantified using ImageJ software. We also checked an eventual difference between the cases when the magnetic field was applied before and after the flow onset. The steady-state size and shape of the clouds were not affected by the sequence of field / flow switching on.

Aqueous solutions of iron oxide nanoparticles (ferrofluids) were synthesized by a coprecipitation of ferrous and ferric salts in an alkali medium and subsequently stabilized by an appropriate amount of oleic acid and sodium oleate using the method described in details by Wooding et al. [33] and Bica et al. [34]. Magnetic nanoparticles were characterized by the transmission electron microscopy (TEM), dynamic light scattering (DLS), ζ-potential measurements and vibrating sample magnetometry (VSM). The characterization results are described in details in [7]. Briefly, TEM pictures and DLS measurements reveal that the iron oxide nanoparticles (of a volume mean diameter of 13 nm) were gathered into irregularly shaped nanoclusters of a mean sphericity close to the unity and of a volume mean diameter equal to 62 nm. Aggregation of nanoparticles occurred during the synthesis likely because of an uncontrolled kinetics of the second surfactant layer adsorption. The first surfactant layer (oleic acid deprotonated in alkali medium) was chemically adsorbed by its $COO^-$ group on the external surface of iron oxide nanoclusters, and the second layer (sodium oleate) was physically adsorbed onto the first one such that its polar $COO^-$ groups pointed outside the nanocluster towards the aqueous solvent. Such a steric double layer, bearing a quite strong negative charge (ζ-potential about -60 mV at a pH=8-9 and ionic strength ranging from 4 to 7 mM), ensured a rather good colloidal stability of synthesized ferrofluids: nanoclusters did not sediment during at least half a year. However, their relatively big size allowed a significant amplification of magnetic interactions and improved substantially their capture efficiency, as compared to single nanoparticles. The initial aqueous suspension contained 4.2 %vol. of nanoclusters and was diluted by a distilled water (milli-Q, 18.2 MΩ·cm) in order to obtain dilute suspensions of the solid phase volume fraction $\phi_0$=0.32 %vol.

Since the nanocluster behavior is principally governed by magnetic interactions, their magnetization properties are of particular importance. They are inspected in more details in Fig 2 where we plot the magnetization curve of the dry powder of iron oxide nanoclusters. This curve has a shape reminiscent for Langevin magnetization law. Saturation magnetization and initial magnetic susceptibility (slope at the origin) are found to be equal to $M_S$=290±10 kA/m and $\chi_i$=9.0±0.5. The latter value allows us to estimate the initial magnetic permeability of the individual nanoclusters, $\mu_n$≈30 – the value given by the model of multipole interactions between nanoclusters [see Section IV-A, Eq. (4)]. As inferred from the inset of Fig. 2, the magnetization curve of the iron oxide powder is nearly linear in the range of the magnetic field intensities $H_0$=0-12 kA/m, used in our experiments. This allows us to suppose that, within the experimental field range, nanocluster magnetic permeability is independent of the applied magnetic field and equal to $\mu_n$≈30.

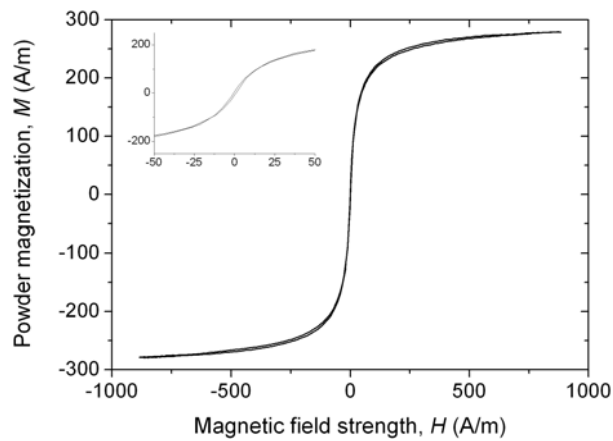

Fig.2. Magnetization curve of the dry powder of iron oxide nanoclusters

We have also checked by magnetization measurements that nickel microparticles preserved their strong magnetic properties after having been heated in the oven following the protocol similar to the one used for their adhesion on the glass plate. The initial magnetic permeability and saturation magnetization of thermally treated nickel particles is estimated to be equal to $\mu_m \sim 10^2$ and $M_{S,m} \approx 450$ kA/m, respectively.

### III. Overview of observation results

A sequence of pictures of the nanocluster clouds accumulated around a nickel microparticle in the presence of a magnetic field (of an intensity $H_0$=12 kA/m) longitudinal to the flow is shown in Fig. 3 for the suspension of the solid phase volume fraction $\phi_0$=3.2·10$^{-3}$ for different suspension flow rates $Q$, corresponding to the superficial velocities $v_0 = Q/S$ ranging from 0 to 1.79·10$^{-3}$ m/s, with $S$ being the flow channel cross-section. As a reference, a bare nickel microparticle in the absence of a magnetic field is shown in Fig. 3a. A picture of the microparticle bearing two nanocluster clouds in the presence of the external field but in the absence of flow is shown in Fig.3b. The applied magnetic field magnetizes the microparticle, and the latter attracts the iron oxide nanoclusters. We were unable to see single nanoclusters because of the optical resolution limit, but observed a change of the suspension optical contrast in the vicinity of the nickel microparticle because of the redistribution of nanocluster concentration. In more details, the nanoclusters accumulate near magnetic poles of the nickel microparticle and are repelled from the equatorial circumference of the microparticle. Such anisotropy of the nanocluster clouds in the absence of flow have been recently observed by Magnet et al. [7] and explained by anisotropy of magnetic interactions favoring attraction within the region where the local magnetic field $H$ is higher than the external field $H_0$ and repulsion within the region where $H<H_0$.

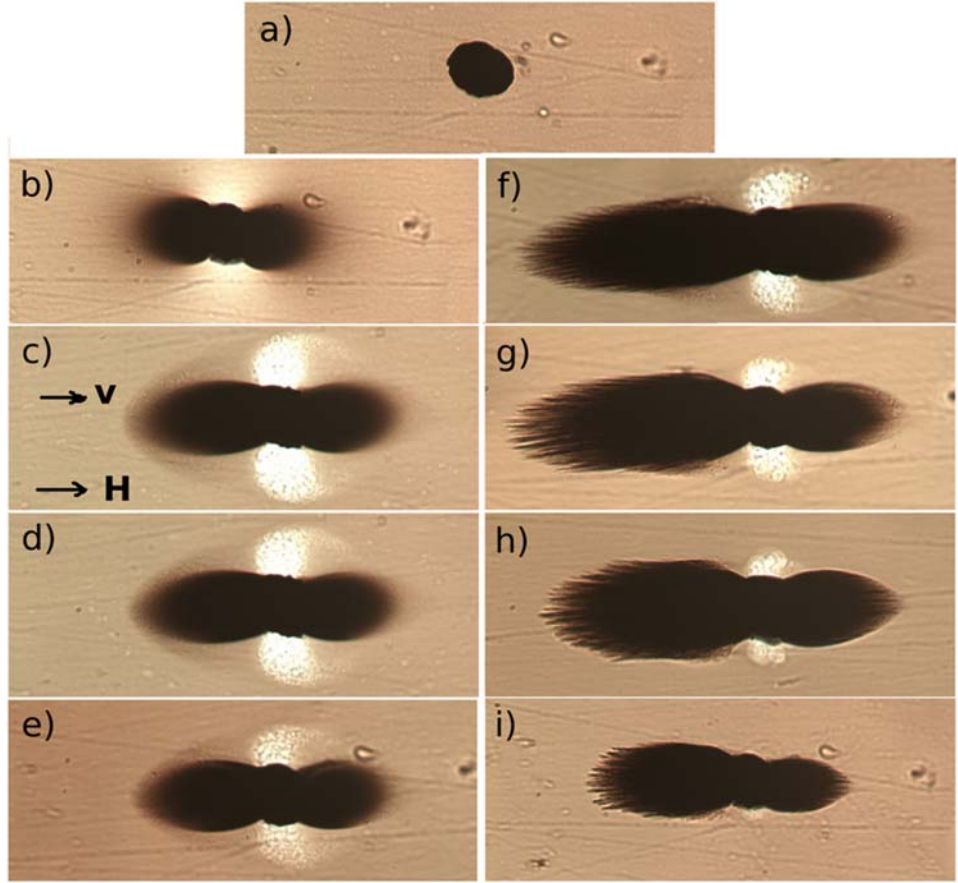

Fig.3. (Color online) Visualization of nanocluster clouds around a nickel microparticle in the longitudinal magnetic field, $H_0$=12 kA/m at the volume fraction of solids in the suspension equal to $\phi_0$=0.32%. Snapshot (a) shows a bare microparticle. Snapshot (b) illustrates nanocluster clouds formed in the absence of flow, but in the presence of an external horizontal magnetic field at the elapsed time equal to ten minutes after the field application. Snapshots (c)-(i) show the nanoclister clouds in the presence a magnetic field and in the presence of the flow oriented from the left to the right of the figure, parallel to the magnetic field direction. These snapshots were taken two hours after the moment of the flow onset. The superficial velocity, $v_0$, of the flow is equal to $1.67 \cdot 10^{-4}$ m/s (c), $2.38 \cdot 10^{-4}$ m/s (d), $3.10 \cdot 10^{-4}$ m/s (e), $4.05 \cdot 10^{-4}$ m/s (f), $5.95 \cdot 10^{-4}$ m/s (g), $1.19 \cdot 10^{-3}$ m/s (h) and $1.79 \cdot 10^{-3}$ m/s (i).

Figures 3 c-i show the cloud shape under flow, two hours after the flow onset, when the steady state regime was achieved. The flow is from the left to the right of the pictures in the same direction as the external magnetic field. We see that the flow induces an asymmetry of the clouds. The front cloud (facing toward the arriving suspension flux) appears to be somewhat larger than the back cloud (situating behind the nickel microparticle) and this difference depends on the suspension velocity. First, at low speeds, $v_0 \leq 3.10 \cdot 10^{-4}$ m/s, the cloud size seems to be almost constant, then it exhibits a step-wise increases at $v_0 = 4.05 \cdot 10^{-4}$ m/s followed by a regular monotonic decrease at higher speeds. A relatively small cloud size at small speeds could be explained as follows. The external magnetic field induces a phase separation in the bulk of the magnetic suspension independently of the presence of nickel microparticles. This phase separation is manifested through the formation of the rod-like aggregates composed of magnetic nanoclusters. The aggregates grow rather quickly thanks to shear-induced collisions and quite strong magnetic interactions between nanoclusters. So, they become visible in optical microcope a few minutes after the field application. On the other hand, they are subjected to gravitational sedimentation because of the density difference with the aqueous solvent. At low suspension speeds their travel time from the channel inlet to the nickel microparticle appears to be larger than the time required for their settling across the channel thickness $h \approx 70$ μm. Because of friction with the channel bottom, the aggregates cannot move once they have been settled. Therefore, the clouds are principally built by the aggregates formed in the vicinity of the microparticle a few moments after the flow onset. At higher speeds, the settling time is longer than the travel time, and the aggregates continuously arrive to the microparticle and form relatively large clouds. Their size and shape achieve steady-state at much longer elapsed times (about one hour) and are defined by the interplay between magnetic and hydrodynamic interactions and, eventually, Brownian motion of nanoclusters, as will be shown in Section IV. With increasing velocity (Figs. 3f-i), hydrodynamic forces become more important, so that a stronger magnetic field is needed to maintain the nanoclusters on the cloud surface. Therefore the part of the cloud situating far from the microparticle is washed away and the cloud surface becomes closer to the microparticle where the magnetic field is high enough to maintain the nanoclusters.

At all suspension velocities, including zero, the nanoclouds are completely opaque. This does not allow us to estimate the nanocluster concentration inside the clouds by the measurements of the transmitted light intensity. Theoretical analysis [7] (see also Section IVA) shows that this concentration is high enough, so that the nanoclusters likely undergo a condensation phase transition at the magnetic field, $H_0=12$ kA/m, used in our experiments. They form solid-like clouds around a microparticle and a dilute fluid-like phase around the clouds. A diffuse border of the clouds at zero and small speeds, $v_0 < 3.10 \cdot 10^{-4}$ m/s [Figs. 3b-e], could be attributed to the polydispersity of the nanocluster suspension. Larger nanoclusters possess a higher magnetic energy and are accumulated in the vicinity of the microparticle forming a dense solid-like phase, while smaller nanoclusters form a diffuse layer around the solid region.

At higher speeds, $v_0 > 4.05 \cdot 10^{-4}$ m/s, the diffuse layer seems to disappear and a smooth shape of the cloud is replaced by a sharp pattern with conical spikes on its surface [Figs. 3f-i]. Similar spikes have been observed in the vicinity of the magnetic poles of concentrated ferrofluid micro-droplets formed in the bulk ferrofluid because of the phase separation [35, 36]. Such a surface instability has been explained in terms of the surface energy anisotropy that favors some surface directions over others. Simulations, assuming arrangement of magnetic particles in body-centered-tetragonal (BCT) lattice, have revealed negative surface energies when the angle, $\delta$, between the surface and the field direction becomes larger than 31deg [37]. The flat surfaces with $\delta > 31$ deg are therefore absolutely unstable, while appearance of spikes with apex angles, $\delta < 31$ deg is energetically favorable. More recently, Cebers [38] has carried out rigorous numerical simulations of the kinetics of the magnetic colloid phase transition and found a multi-spike shape of the droplets of the concentrated colloid phase attributing it to the surface tension anisotropy.

Formation of spikes is expected at any applied magnetic field strong enough to induce a solid-fluid phase separation. However, in our case, it is not observed in the absence of flow, neither at low speeds [Figs. 3b-e], provided that the flow should not affect significantly the balance of the surface stresses, according to the estimations presented in Section IV B. The absence of peaks is likely connected to the diffuse boundary layer, which probably destroys the source of the surface instability – the negative surface energy. On the contrary, disappearance of the diffuse layer at speeds $v_0 \geq 4.05 \cdot 10^{-4}$ m/s leads to a sharp interface of the clouds with an appropriate surface energy and may induce instabilities. Disappearance of this layer likely comes from hydrodynamic forces, which squeeze small nanoclusters to the solid-like phase of the clouds or wash them away from clouds. Finally note that the above considered surface instability appears in the case when the surface energy is principally governed by the magnetic interactions between the particles belonging to the surface layer – the case of the interface between two different phases of the same magnetic suspension. This instability should not be confused with the Rosensweig instability [39] of the interface between two distinct magnetic suspensions (ferrofluids) subjected to an orthogonal magnetic field and whose surface energy is governed by molecular interactions and considered to be field-independent.

A video of the flow around the microparticle with attached nanocluster clouds corresponding to the picture shown in Fig. 3i (at $H_0$=12 kA/m and $v_0$=1.79·10$^{-3}$ m/s) is presented in Supplemental material [40]. As inferred from this video, the water flux arriving on the front cloud makes the spikes moving along the surface in the direction of the streamlines. The rear part of the back cloud has a tapered shape favorable to the flow. Such a shape likely induces only minor perturbations of the flow so that the spikes on the rear cloud seem to be quasi-immobile. We also observe some recirculation of nanocluster aggregates in the vicinity of the points where the cloud surface joins the microparticle. This recirculation is more pronounced on the lower side likely because of an imperfect alignment between the field and the flow.

Another point revealed in visualization experiments is appearance of bright white regions near the equatorial circumference of the microparticle. These regions correspond to aggregates of micelles of non-adsorbed oleic acid. As already stated, strongly magnetic nanoclusters are expelled from the equatorial region while nonmagnetic oleic acid aggregates are forced to move there because of the volume conservation of the whole suspension. The quantity of the captured oleic acid decreases with increasing flow speed, because hydrodynamic forces become more important as compared to the effective attraction.

We should also mention that the accumulation of nanocluster clouds around a magnetized microparticle is completely reversible process: once the magnetic field is switched off, the cloud is completely dissolved by Brownian motion and by the water flux streaming the microparticle. Destruction of the clouds after switching off of the magnetic field, of an intensity $H_0$=12 kA/m, is demonstrated in the second video posted in Supplemental Material [41]. The reversibility of the cloud formation / dissociation could be explained by the absence of remnant magnetization of the nanoclusters [Fig. 2] and by a presumably low solid friction between the nanoclusters covered by a surfactant double layer.

A sequence of pictures showing a steady-state shape and size of nanocluster clouds in the presence of a magnetic field perpendicular to the flow is shown in Fig. 4 for $H_0$=12 kA/m, the suspension volume fraction, $\phi_0$=0.32%, and flow velocities ranging from $v_0$=1.67·10$^{-4}$ to 2.38·10$^{-3}$ m/s. The shape of the clouds appears to be quite similar to the one in the longitudinal field. The clouds are extended along the applied magnetic field and conical spikes appear on their extremities because of the surface energy anisotropy. Both clouds attached to the microparticle have the same size and shape because of the symmetry of the streamlines with respect to the plain perpendicular to the applied magnetic field and passing through the microparticle center. The clouds seem however to be slightly asymmetric with respect to the microparticle axis aligned with the field. Such an asymmetry is caused by the hydrodynamic drag pushing the clouds in the direction of the flow. As inferred from Fig. 4, the cloud size decreases progressively with an increasing flow speed, $v_0$, that is explained in terms of increasing hydrodynamic forces washing the nanoclusters away from the clouds.

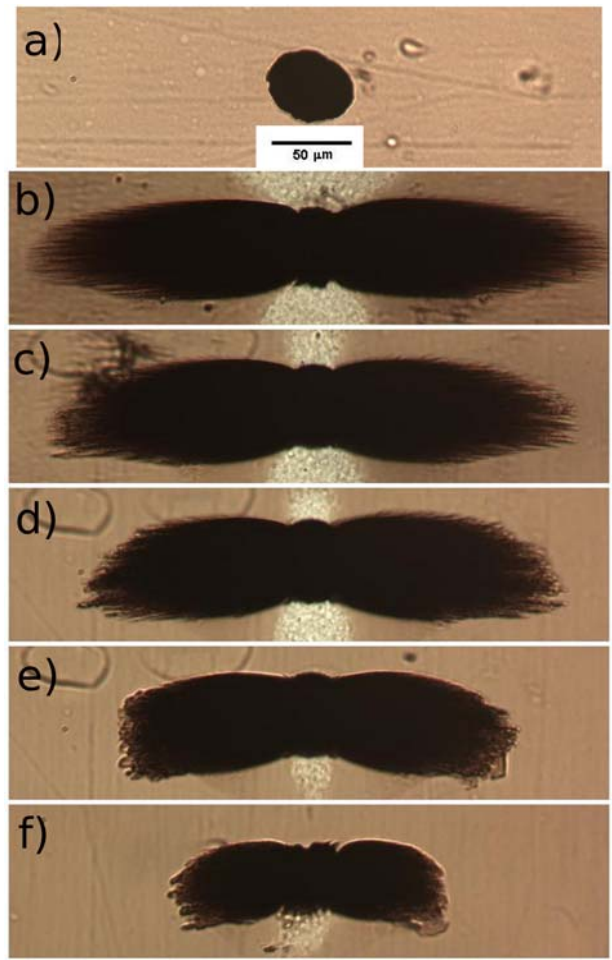

Fig.4. (Color online) Visualization of the nanocluster clouds in the transverse magnetic field $H_0$=12 kA/m at different flow speeds $v_0$, equal to 1.67·10$^{-4}$ m/s (b), 2.38·10$^{-4}$ m/s (c), 5.95·10$^{-4}$ m/s (d), 1.19·10$^{-3}$ m/s (e) and 2.38·10$^{-3}$ m/s. The snapshot (a) shows a bare nickel microparticle.

The mechanisms defining the cloud size and shape will be inspected in the next Section IV where experimental results will be compared with predictions of our model.

IV. Theory and discussion

The field-induced condensation phase transition is a distinguishing feature of our system having a strong impact on the nanocluster accumulation around a magnetized microparticle. Therefore, we begin with a thermodynamic description of this phase transition in the absence of flow (Sec. IV A). In the presence of flow, the cloud behavior, size and shape depend on Brownian motion, magnetic and hydrodynamic forces acting on nanoclusters. Simultaneous consideration of these three effects along with the condensation phase transition would substantially complicate the theoretical description. Fortunately, the thermodynamics governing the phase transition and the hydrodynamics defining the cloud size can be decoupled for relatively low suspension speeds, considered in our experiments. The validity of such decoupling is proved in Sec. IV B where we estimate the relative importance of hydrodynamic and magnetic forces, or rather their ratio, called Mason number. Based on this estimation, we calculate the cloud shape (Sec. IV C) and size (Sec. IV D) under the field and the flow in the steady-state regime. We consider the case of the field parallel to the flow. Finally, we compare the calculated cloud size to the one observed in experiments (Sec. IV E).

A. Phase transition

The appropriate parameter describing relative importance of magnetic interactions is the so-called dipolar coupling parameter. It is defined as the ratio of magnetic-to-thermal energy ($kT$) of the nanocluster, and scales as:

$$\alpha = \frac{\mu_0 \beta_n^2 H_0^2 V_n}{2kT}, \quad (1)$$

where $\mu_0 = 4\pi \cdot 10^{-7}$ H/m is the magnetic permeability of vacuum, $V_n$ is the nanocluster volume, $\beta_n = (\mu_n - 1)/(\mu_n + 2)$ is the magnetic contract factor of the nanocluster and $\mu_n$ is the nanocluster magnetic permeability. The factor $\beta_n^2$ in the last equation comes from the energy of dipole-dipole interaction between magnetic nanoclusters proportional to the square of their magnetic moment.

The dipolar coupling parameter $\alpha$ is estimated to be of the order of 2 for the experimental value of the magnetic field intensity $H_0 = 12$ kA/m. However, such relatively modest value of this parameter appears to be sufficient to induce a phase separation in the suspension of magnetic nanoclusters of the magnetic permeability as high as $\mu_n \approx 30$. Since the magnetic field and the nanocluster concentration are not homogeneous around the microparticle, we should check the phase behavior of the ensemble of nanoclusters at different concentrations and applied magnetic fields. To this purpose, we shall construct a phase diagram, $\alpha$ - $\Phi$, where different phases will be identified. A similar phase diagram has already been developed via Monte-Carlo simulations or analytical calculations for the magnetic particles exhibiting dipole-dipole interactions [42, 43]. In our case of magnetic nanoclusters with a strong magnetic permeability, $\mu_n \approx 30$, we should take into account short-range multipolar interactions, which are especially important at moderate-to-high concentrations, since the dipolar interactions strongly underestimate the strength of the interactions between particles [44].

To proceed with, we assume that all the nanoclusters are identical and only two phases of the nanocluster ensemble may exist: a disordered fluid and an ordered solid having a face-centered cubic (FCC) structure. Even though the body-centered tetragonal (BCT) lattice has the least energy in the presence of reasonably high magnetic fields [45], our choice for the FCC lattice is motivated by the desire to capture the disorder-order phase transition at zero field keeping in mind that that the energy of both structures differs insignificantly. Neglect of other possible ordered states should not cause substantial errors in determination of the nanocluster concentration profile around a magnetized microparticle. The equilibrium between the two considered phases is found by the equilibrium of nanocluster chemical potentials, $\zeta$, and osmotic pressures, $p$, in each phase [46]:

$$\zeta(\Phi_s, \alpha) = \zeta(\Phi_f, \alpha) \qquad (2a)$$

$$p(\Phi_s, \alpha) = p(\Phi_f, \alpha) \qquad (2b)$$

where $\Phi$ is the nanocluster concentration in the suspension and the subscripts "$s$" and "$f$" stand for the solid and fluid phases, respectively. Taking into account the porous nature of the nanoclusters, their concentration is related to the "true" volume fraction of solids, $\phi$, by the following relation: $\Phi = \phi / \Phi_n$, with $\Phi_n \sim 0.5$ being the internal volume fraction of the nanoclusters.

The chemical potential $\zeta(\Phi, \alpha)$ and the osmotic pressure $p(\Phi, \alpha)$, both contain the contributions coming from magnetic interactions (considered in details in [7]) and hard-sphere repulsion. The appropriate expressions for the later interaction have been developed by Zubarev and Iskakova [47] on the basis of osmotic compressibility calculations carried out by Carnahan and Starling [48] and Hall [49] for semi-dilute and concentrated hard-sphere suspensions. Assuming that the steric interactions between our nanoclusters respect the Carnahan-Starling law in the fluid phase and the Hall law in the solid phase, we arrive to the following expressions for quantities $\zeta$ and $p$ in both phases:

$$\zeta(\Phi_f, \alpha) = (\zeta_{hard-sphere})_f + (\zeta_{magn})_f = kT \left( \ln \Phi_f + \Phi_f \frac{8 - 9\Phi_f + 3\Phi_f^2}{(1-\Phi_f)^3} \right) - kT \frac{\alpha}{\beta_n^2} \frac{\partial \mu}{\partial \Phi_f} \qquad (3a)$$

$$p(\Phi_f, \alpha) = (p_{hard-sphere})_f + (p_{magn})_f = \frac{kT}{V_n} \Phi_f \frac{1 + \Phi_f + \Phi_f^2 + \Phi_f^3}{(1-\Phi_f)^3} - \frac{kT}{V_n} \frac{\alpha}{\beta_n^2} \left( \Phi_f \frac{\partial \mu}{\partial \Phi_f} - \mu + 1 \right)$$

(3b)

$$\zeta(\Phi_s, \alpha) = (\zeta_{hard-sphere})_s + (\zeta_{magn})_s = kT \left( \frac{A}{\Phi_m} \ln \frac{\Phi_s}{\Phi_m - \Phi_s} + \frac{A}{\Phi_m - \Phi_s} + C \right) - kT \frac{\alpha}{\beta_n^2} \frac{\partial \mu}{\partial \Phi_s} \qquad (3c)$$

$$p(\Phi_s,\alpha) = (p_{hard-sphere})_s + (p_{magn})_s = \frac{kT}{V_n}\Phi_s\frac{A}{\Phi_m - \Phi_s} - \frac{kT}{V_n}\frac{\alpha}{\beta_n^2}\left(\Phi_s\frac{\partial \mu}{\partial \Phi_s} - \mu + 1\right) \qquad (3d)$$

where $\Phi_m = \pi/(3\sqrt{2}) \approx 0.74$ is the maximum packing fraction of the FCC structure, $A \approx 2.2$ and $C \approx 1.255$ are the constants ensuring order-disorder phase transition at zero field in the concentration range $0.495 < \Phi < 0.545$ [50]. The magnetic permeability of the nanocluster suspension, $\mu$, intervening into the last equations is found assuming multipolar interactions between magnetic nanoclusters arranged in the FCC lattice. The interparticle distance in the three directions of the lattice is imposed by the suspension volume fraction. Performing numerical simulations using the numerical code developed by Clercx and Bossis [44], we obtain the following interpolating function for the magnetic permeability as function of the nanocluster concentration, $\Phi$, and nanocluster magnetic permeability, $\mu_n$:

$$\mu(\Phi,\mu_n) = (\mu_{MG})^{\exp\left[(c_1 \log^2 \mu_c + c_2 \log \mu_c)\Phi^6\right]} \qquad (4)$$

where $\mu_{MG} = (1 + 2\beta_n\Phi)/(1 - \beta_n\Phi)$ is the dilute-limit expression of the suspension magnetic permeability given by the Maxwell-Garnett mean field approach [51], $c_1 \approx 0.408$ and $c_2 \approx 0.12$ are numerical constants.

The magnetic permeability of the nanoclusters, $\mu_n$, can be found from the experimental magnetization curve of the dry powder of nanoclusters [Fig.2]. The slope at the origin of this curve gives the initial magnetic susceptibility of the powder, $\chi_p \approx 9$, which corresponds to the magnetic permeability $\mu_p = \chi_p + 1 \approx 10$. Assuming that the concentration dependency of the powder magnetic permeability is similar to that of the liquid suspension, i.e. defined by Eq.(4), we solve this equation with respect to $\mu_n$ having replaced $\mu(\Phi, \mu_n)$ by $\mu_p \approx 10$ and $\Phi$ by $\Phi_m \approx 0.74$. In this way, we obtain an estimation for the nanocluster magnetic permeability, $\mu_n \approx 30$, at relatively low magnetic fields, $H \leq 12$ kA/m, considered in the present work.

Having defined all the terms intervening into Eqs. (2a), (2b), we obtain a system of two transcendental equations, which is solved with respect to $\Phi_s$ and $\Phi_f$ for given values of the dipolar coupling parameter $\alpha$. So obtained functions, $\Phi_s(\alpha)$ and $\Phi_f(\alpha)$ correspond to the binodal curves of the $\alpha$-$\Phi$ diagram plotted in Fig. 5. These two curves separate the phase diagram into the three regions, as follows: the disordered fluid situating below the left binodal curve; the FCC-solid situating below the right binodal curve and the fluid-solid mixture occupying the space between the two binodals. As expected, at zero applied field ($\alpha=0$) we recover the order-disorder transition in the well known concentration range, $0.495 < \Phi < 0.545$ [50]. At the field parameter $\alpha > 2$, the solid phase exist only in a narrow range of concentrations, whose values are very close to the maximum packing fraction of the FCC lattice, $\Phi_m \approx 0.74$, at least in thermodynamic equilibrium, i.e. in the absence of flow and at a long elapsed time after the moment of the field application.

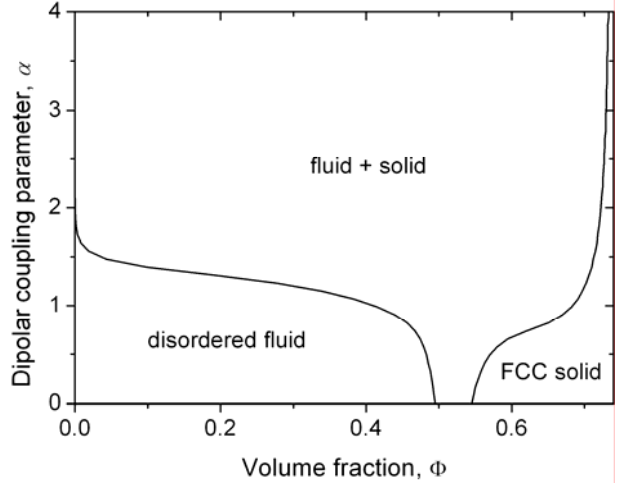

Fig.5. Phase diagram of the suspension of magnetic nanoclusters having a constant magnetic permeability equal to $\mu_n$=30.

**B. Mason number**

In what follows, we shall demonstrate that the cloud internal structure should not change drastically in the presence of flow at the flow speeds used in our experiments. To this purpose, we estimate the characteristic ratio $\sigma_h/\sigma_m$ of hydrodynamic to magnetic stresses acting inside the solid-like cloud. These both stresses scale as $\sigma_h \sim \eta_i v_i / r_m \sim \eta_0 v_0 / r_m$ and $\sigma_m \sim \mu_0 H_0^2$. Here $\eta_0$ and $\eta_i$ are the viscosities of the suspending liquid (water) and of the particle suspension inside the cloud, $v_0$ and $v_i$ are the suspension velocities far upstream from the microparticle and inside the cloud, respectively, $r_m$ is the microparticle radius. It is easy to show that the product $\eta_i v_i$ of the suspension viscosity by its characteristic speed inside the clouds appears to be of the same order of magnitude as $\eta_0 v_0$ if one assumes a recirculation flow inside the clouds in the limit of high particle concentration ($\Phi \sim \Phi_m \approx 0.74$ resulting in $\eta_i \gg \eta_0$). The ratio of stresses therefore reads:

$$\frac{\sigma_h}{\sigma_m} \sim \frac{\eta_0 v_0}{\mu_0 H_0^2 r_m} \qquad (5)$$

Estimations show that this ratio is quite low in our experimental conditions: $10^{-5} < \sigma_h/\sigma_m < 10^{-3}$. Small values of $\sigma_h/\sigma_m$ (responsible for the internal flow and for the cloud surface behavior) allow us to conclude that the flow should have a minor effect on the cloud internal structure and on the shape of its surface. This will make possible to determine the internal volume fraction and the shape of the clouds in the limit of zero Mason number, in a similar way as in the absence of flow.

On the other hand, the cloud volume is governed by the flux of nanoclusters arriving on the clouds and leaving the clouds. The key parameter is the ratio of the convective to the magnetic migration flux, or rather the ratio of the hydrodynamic-to-magnetic forces acting on the nanoclusters in the vicinity of the cloud surface, so called Mason number:

$Ma = F_h/F_m \sim (\eta_0 v r_n)/(\mu_0 m |\nabla H|)$. Here $r_n$ is the nanocluster radius, $m \sim \beta_n H_0 r_n^3$ is the nanocluster magnetic moment, $|\nabla H| \sim \beta_m H_0/r_m$ is the magnetic field gradient induced by the microparticle, $\beta_m = (\mu_m - 1)/(\mu_m + 2)$ is the magnetic contract factor of the microparticle and $\mu_m \sim 10^2$ is the microparticle magnetic permeability. The flow speed, $v$, outside the cloud (at a distance of the order of one nanocluster radius, $r_n$, from the cloud surface) scales as: $v \sim \dot{\gamma} r_n \sim v_0 (r_n/r_m)$, with $\dot{\gamma} \sim v_0/r_m$ being the shear rate at the cloud surface. Thus, the Mason number scales as:

$$Ma = \frac{\eta_0 v_0}{\mu_0 \beta_m \beta_n H_0^2 r_n} \qquad (6)$$

Note that this definition of the Mason number is somewhat more general than the one conventionally used in electro- and magnetorheology [52], where the attraction force between two identical particles is considered. In this particular case, $\beta_m = \beta_n = \beta$ and $\beta^2$ would appear in Eq. (6) instead of the product $\beta_m \beta_n$ of magnetic contrast factors of both particles. Estimations give moderate values of the Mason number for our experimental parameters: $10^{-2} < Ma < 1$. This suggests that the nanocluster motion outside the clouds should be affected by the flow. Thus, the nanoclusters may be washed away from the cloud, which would limit the cloud growth; the cloud size is therefore expected to be governed by the Mason number $Ma$.

### C. Cloud shape

Now, we are ready to calculate the cloud shape and size in the longitudinal magnetic field taking into account the above estimations. Firstly, based on the phase diagram [Fig. 5] and neglecting small hydrodynamic stresses [Eq. (5)], we shall consider that the nanocluster concentration is homogeneous inside the cloud and equal to the maximum packing fraction of the FCC lattice, $\Phi_m \approx 0.74$. The magnetic permeability of the quasi-solid cloud is calculated by Eq. (4) and is equal to $\mu \approx 10$. The nanocluster concentration and magnetic permeability outside the cloud will be neglected. Secondly, we remark that the magnetic permeability of nickel microparticles, $\mu_m \sim 10^2$, is an order of magnitude higher than that of the cloud. We assume therefore that the magnetic field distribution inside the cloud is the same as the one around an isolated magnetized sphere of an infinite magnetic permeability placed into a non-magnetic medium. This approximation is not really justified, nevertheless, it allows obtaining a correct shape of the clouds avoiding substantial numerical efforts. Note that this approximation does not introduce the field distribution outside the cloud. Thus, it does not contradict to the discontinuity of the magnetic field and of the pressure at the cloud surface. This discontinuity will be taken into account. Thirdly, introduce the polar coordinate system ($r$, $\theta$), as shown in Fig. 6 with the origin (point O) at the microparticle center, the polar angle $\theta$ counted in the clock-wise direction from the Oz parallel to the lines of the external field. Let all the distances be dimensionless and normalized by the microparticle radius, $r_m$.

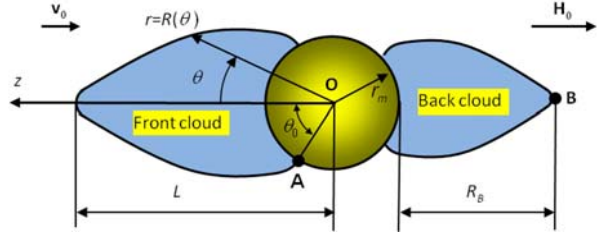

Fig. 6. (Color online) Sketch of the problem geometry

At this stage we do not intent to describe precise morphology of the cloud surface with eventual appearance of the conical spikes. Instead of this, we try to capture a global shape of the cloud, which will allow us to correctly estimate its size. Therefore, the cloud surface is supposed to be axisymmetric and smooth. The cloud surface is described by a geometric locus $[R(\theta), \theta]$ in polar coordinates. The function $R(\theta)$ is found from the balance of normal stresses on the cloud surface, where we may neglect the viscous stress, $\sigma_h$, because it is a few orders of magnitude smaller than the magnetic stress, $\sigma_m$ [c.f. Eq. (5)]. Under this condition, the continuity of the normal stresses on the cloud surface will give us the discontinuity of the pressure on the cloud surface, which reads: $p_o - p_i = \mu_0 M_n^2 / 2$, where the subscripts "$o$" and "$i$" stand for the outer and inner faces of the cloud surface and the right-hand side of this equation represents the magnetic pressure jump proportional to the square of the component $M_n$ of the suspension magnetization normal to the cloud surface. Neglecting the capillary pressure jump, in our previous work, we have shown that the pressure balance on the cloud surface reduces to the following differential equation with respect to the desired function $R(\theta)$ [7]:

$$h_0^2 - h^2 = (\mu - 1)\frac{(h_r - h_\theta R'/R)^2}{1 + (R'/R)^2} \qquad (7)$$

where $R' = dR/d\theta$; $h_r = \left(1 + 2\beta_m / R^3\right)\cos\theta$ and $h_\theta = -\left(1 - \beta_m / R^3\right)\sin\theta$ are respectively, the radial and the polar components of the magnetic field intensity, both normalized by the applied field $H_0$, $h = \sqrt{h_r^2 + h_\theta^2}$ is absolute value of the normalized magnetic field at a given point $(R, \theta)$ of the cloud and $h_0 \approx 3\cos\theta_0$ (with $\beta_m \approx 1$ at $\mu_m \sim 10^2$) stands for the value of $h$ at the point ($R=1$ and $\theta=\theta_0$) where the cloud surface joins the microparticle; this point will be hereinafter referred to as the anchoring point (point A in Fig. 6) and the angle $\theta_0$ will be called the anchoring angle. The ordinary differential equation (7) is solved numerically at the initial condition $R(\theta_0)=1$.

In the present model, the anchoring angle defines the cloud volume. Varying this angle in an artificial manner from 0 to a critical value $\theta_0 \approx 1.2$ (or 69 deg), we see that the cloud volume progressively increases. At the angles $\theta_0 > 69$ deg, the Eq. (7) does not have a solution in the domain of real numbers because its left hand side turns to be negative while the right hand side is always positive. This means that the above considered assumptions (2D-axisymmetric shape without spikes and capillary effects) cannot ensure existence of clouds with anchoring angles above 69 deg. At the same time, image processing of the experimental snapshots reveal the angles lower than or equal to 69±2 deg for all experimental parameters considered in the present work. This allows us to apply the present model with confidence for estimations of the cloud shape.

According to the arguments presented in the Sec. IV-B, the cloud volume is affected by the flow and should depend on the Mason number *Ma*. It is therefore indispensible to find the influence of the flow on the anchoring angle related to the cloud size.

### D. Anchoring angle: effect of the flow

The starting point is that the amount of nanoclusters inside the clouds does not change during time in the steady-state regime. This implies that the total flux of nanoclusters across the cloud surface, *S*, is zero for both clouds attached to the microparticle: $J = \iint_S \mathbf{j} \cdot \mathbf{n} dS = 0$, with **j** being the flux density vector and **n** – the outward normal unit vector at the cloud surface.

For the better understanding, let us consider the cloud behavior in more details. Under Brownian motion, the nanoclusters situated at the cloud surface are subjected to the diffusion: they tend to «evaporate» from the surface and to displace towards the regions of weak concentration. In the absence of flow, the diffusive flux density $\mathbf{j}_d$ is totally counterbalanced by the so-called magnetophoretic flux density $\mathbf{j}_m$ (responsible for particle migration along the magnetic field gradient). This latter flux causes the «evaporated» nanoclusters to go back to the clouds, such that the total flux density is zero:

$$\mathbf{j}_{eq} = \mathbf{j}_{d,eq} + \mathbf{j}_{m,eq} = \mathbf{0} \tag{8a}$$

where the subscript "*eq*" stands for the flux magnitudes at the thermodynamic equilibrium. The flow modifies this equilibrium of fluxes adding a convective flux associated to the fluid velocity **v**:

$$\mathbf{j} = \mathbf{j}_d + \mathbf{j}_m + \Phi \mathbf{v} \tag{8b}$$

where $\Phi$ is the volume fraction of nanoclusters on the outer side of the cloud surface.

The magnetic force acting on nanoclusters at the cloud surface increases when the cloud volume decreases and its surface becomes closer to the microsphere. Thus, at small cloud volumes, the magnetic force is relatively large, and the hydrodynamic force is insufficient to wash the nanoclusters away from the cloud. The cloud grows during time. The situation becomes opposite at large cloud volumes when the cloud size is forced to decrease under erosive hydrodynamic forces. In both cases of large or small clouds, their volume tends to the same final value, corresponding to the steady-state. This clearly shows that the cloud size is governed by the particle flux balance (8b), as well as by the interplay between the hydrodynamic and the magnetic forces, which is described by the Mason number *Ma* [Eq. (6)].

Precise determination of the flux density **j** requires solution of the free surface boundary value problem connected with coupled convective-diffusion, Stokes and Maxwell equations. To obtain a reasonable estimation of the cloud size avoiding substantial numerical efforts, we shall introduce some simplifying assumptions.

First, our experiments reveal some parts of the cloud surface where the nanoclusters seem to neither arrive, nor detach from these parts, implying a zero normal component of the flux density: $j_n \approx 0$. In the longitudinal magnetic field, such zero-flux zone appears on the front cloud in the vicinity of the microparticle, while it seems to extend over the whole surface of the rear cloud [cf. Fig. 3f-i]. These zones were more easily observed in videos.

Secondly, we shall impose the above zero flux-condition, $j_n \approx 0$, at the points of the cloud surface where the tensile hydrodynamic force acting on nanoclusters (and eroding them from the cloud surface) is the highest. These points are the anchoring point A of the front cloud and the rear point B of the back cloud (cf. Fig.6).

Thirdly, we consider a relatively slow flow ($Ma<1$), and suppose that the diffusive flux density $\mathbf{j}_d$ in both considered points A and B is not very different from the one in equilibrium, $\mathbf{j}_{d,eq}$, i.e. in the absence of flow. Strictly speaking, the neglect of the difference of the diffusive fluxes cannot be justified. In theory, this difference can appear to be of the same order of magnitude that the difference of the magnetophoretic fluxes. Nevertheless, quite reasonable agreement of the theoretical results, based on this approximation, with the experimental ones [Sec. IV E] could justify the used simplification $\mathbf{j}_d \approx \mathbf{j}_{d,eq}$, which avoids numerical complexities related to the solution of the free boundary convective-diffusion problem. Thus, eliminating diffusive fluxes from Eqs. (8a) and (8b), we obtain the normal component of the total flux density in the characteristic points A and B:

$$j_n \sim (j_m)_n - (j_{m,eq})_n + \Phi v_n = \Phi \frac{(F_m)_n - (F_{m,eq})_n}{6\pi\eta_0 r_n} + \Phi v_n \approx 0 \qquad (9)$$

where $v_n$ is the component of the suspending liquid velocity normal to the cloud surface and calculated in the vicinity of the outer face of the cloud surface (cf. Appendix). This equation shows that in the points A and B, the normal component of the flux density is proportional to the difference of the normal components of the magnetic forces $(F_m)_n - (F_{m,eq})_n$ exerted on the nanoclusters under flow and in the absence of flow, respectively. The denominator of the first term in the right-hand side of Eq. (9) corresponds to the hydrodynamic friction coefficient of spherical nanoclusters. At the considered approximation, the concentration $\Phi$ works out from the Eq. (9), and this equation reduces to the following force balance at the characteristic points A and B:

$$(F_m)_n - (F_{m,eq})_n \approx -6\pi\eta_0 r_n v_n \qquad (10)$$

This equation shows that the hydrodynamic force compensates the increment of the magnetic force acting on nanoclusters at points A and B of the cloud surface when this surface is displaced by the flow at some distance from its equilibrium position. It should be stressed that this equation applies separately to the points A and B and defines the size of the front cloud when applied to the point A and the size of the back cloud when applied to the point B. More precisely, the equation (10) allows an estimation of the anchoring angle $\theta_0$ as function of the flow velocity. The calculations are developed in details in Appendix. The following transcendental equations are obtained for the anchoring angle of the front (Eq. 11a) and the back (Eq. 11b) clouds:

$$\frac{\cos\theta_0 \sin\theta_0 - \cos\theta_{eq} \sin\theta_{eq}}{\sin\theta_0} \approx \frac{3}{4(2+\beta_m)} Ma \qquad (11a)$$

$$\left(1 + \frac{2\beta_m}{R_B^3(\theta_0)}\right)^2 - \left(1 + \frac{2\beta_m}{R_B^3(\theta_{eq})}\right)^2 = \frac{1}{4} c \cdot Ma \qquad (11b)$$

where $\theta_{eq}$ is the anchoring angle in the absence of flow. It is taken to be equal to the value $\theta_{eq} \approx 69°$ of the critical angle above which our model (Eq. 7) does not provide a stable solution for the cloud shape at equilibrium. $R_B(\theta_0)$ and $R_B(\theta_{eq})$ are the distances between the microsphere center and the point B on the extremity of the back cloud under flow ($Ma>0$) and in the absence of flow ($Ma=0$), respectively [Fig. 6]. The distances $R_B(\theta_{eq})$ and $R_B(\theta_0)$ are obtained by numerical solution of Eq. (7) at the initial conditions $R(\theta=\theta_{eq})=1$ and $R(\theta=\theta_0)=1$, respectively. A numerical factor $c$ has been introduced in Eq. (11b) as an adjustable parameter keeping in mind that we got expressions for the magnetic and hydrodynamic forces, intervening into Eq. (10) up to a dimensionless multiplier [cf. Appendix].

Equations (11a) and (11b) are solved with respect to the angle $\theta_0$ at different values of the Mason number in the range $0<Ma<0.5$, covered in experiments. Theoretical dependencies of the anchoring angle $\theta_0$ on the Mason number are shown in Fig.7 for both the front and the rear clouds. As expected, the anchoring angle of both clouds decreases with the increasing Mason number. This corresponds to a decrease of the cloud volume because of the increase of the convective flux blowing the nanoclusters away from the clouds.

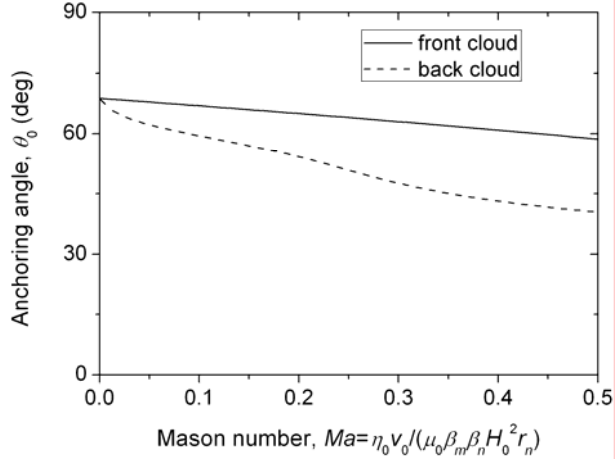

Fig. 7. Theoretical dependencies of the anchoring angle on the Mason number, cf. Eqs. (11a) and (11b) for the front and the back clouds, respectively. The free parameter $c$ in Eq. (11b) is taken to be $c=1.6$ – the value providing the best fit of the theory to the experimental cloud size presented below in Fig. 9

### E. Numerical results and comparison with experiments

The cloud shape, described by the function $R(\theta)$, is found from the solution of the differential equation (7) using the appropriate anchoring angle $\theta_0(Ma)$ at the initial condition $R(\theta_0)=1$. The calculated cloud shape is compared to the experimentally observed one (taken from the snapshots of Fig. 3) in Fig. 8. The cloud shape in the absence of flow is shown in Fig. 8a. As already mentioned, settling of particle aggregates does not allow them to reach the clouds and to establish thermodynamic equilibrium. In experiments, this leads to a much smaller cloud volume than the one calculated by our theory neglecting the settling problem (red solid line). A better agreement (blue dashed line) with experiments is obtained if the calculations are made at the anchoring angle equal to the experimental one, $\theta_0 \approx 57$ deg instead of the equilibrium value $\theta_{eq} \approx 69$ deg. The figures 8b and 8c illustrate the clouds in the longitudinal magnetic field at Mason numbers equal to 0.08 and 0.35 respectively, with the corresponding anchoring angle found from Eqs. (11a) and (11b). As inferred from these figures, our model reproduces qualitatively the elongated shape of the clouds. It correctly predicts an asymmetry of the front and rear clouds in the longitudinal field, as observed in experiments. This asymmetry could come from stronger hydrodynamic forces acting on nanoclusters on the surface of the back cloud (near its rear point B) as compared to the forces near the anchoring point A of the front cloud. Because of the simplifying input hypotheses, the model is unable to predict conical spikes on the cloud surface. Neglecting the capillary pressure, it gives the tapered cloud extremities with singular end points, which is inconsistent with globally rounded shape with numerous spikes observed in experiments. Recall that this surface instability appears as a result of the surface energy anisotropy, as discussed in details in Sec. III.

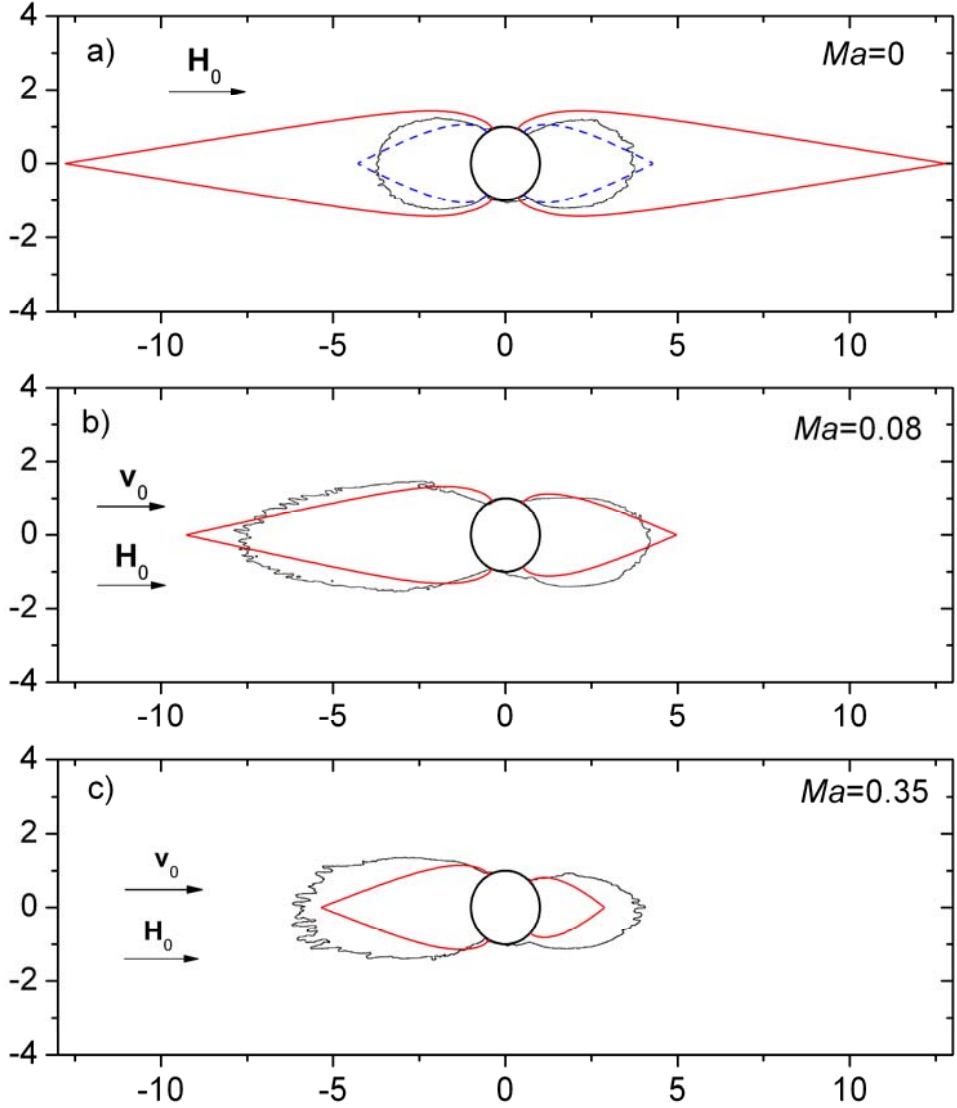

Fig. 8. (Color online) Comparison between the calculated (red solid line) and the experimental (black wavy line) shapes of the nanocluster clouds attached to a microparticle in the presence of an external magnetic field oriented horizontally. The figure (a) shows the clouds in the absence of flow, at $Ma=0$. The blue dashed line on figure (a) corresponds to the calculated shape at the same anchoring angle $\theta_0$ as the one observed in experiments at $Ma=0$. The figures (b) and (c) show the clouds under the flow realized from the left to the right in the direction of the applied magnetic field at Mason numbers $Ma$ equal to 0.08 and 0.35, respectively. The free parameter $c$, intervening into Eq. (11b) is equal to $c=1.6$ and provides the best fit with experimental results on the size of the back cloud [c.f. Fig. 9]. The experimental shape of the clouds has been taken from the snapshots of figure 3: Fig. 3b for (a); Fig. 3f for (b), and Fig. 3i for (c).

To analyze the effect of the flow on the cloud size, we introduce the dimensionless cloud length, $L = R(\theta = 0) - 1$, corresponding to a distance between the microparticle surface and the cloud extremity normalized by the microparticle radius $r_m$ [Fig. 6]. Theoretical and experimental dependencies of the cloud length on the Mason number are shown in Fig. 9 for both orientations of the magnetic field. Experimental data at low Mason numbers $Ma<0.06$, at which the nanocluster aggregates settled before arriving to the microparticles [cf. Figs. 3c-e], have been excluded from this figure. Thus, at the considered range of Mason numbers, $0.06<Ma<0.5$, the length of the front cloud monotonically decreases with the Mason number, which is explained by increasing hydrodynamic forces washing away the nanoclusters from the cloud surface. On the contrary, the length of the back cloud is much less affected by the flow and remains almost constant within the considered range of Mason numbers. Our simple model qualitatively reproduces the decreasing trend for the front cloud and gives a quantitative correspondence with experiments for the longitudinal magnetic field with maximum of 35% of discrepancy. The model predicts a slight decrease of the back cloud size which is not really distinguishable in experiments. It is likely that the rough estimations of the back cloud size, based on the scaling arguments, overestimate the role of the Mason number while, in reality, the back clouds are less sensitive to the flow variations. Recall that the theory does not have adjustable parameters for the predictions of the front cloud size and has a one adjustable parameter for the back cloud. This parameter, $c$, intervenes into Eq. (11b) and gives the best fit to experimental data at $c \approx 1.6$. The experimental cloud length in the transverse field appears to be somewhat smaller than that of the front cloud likely because stronger hydrodynamic forces are exerted to the clouds oriented perpendicularly to the flow.

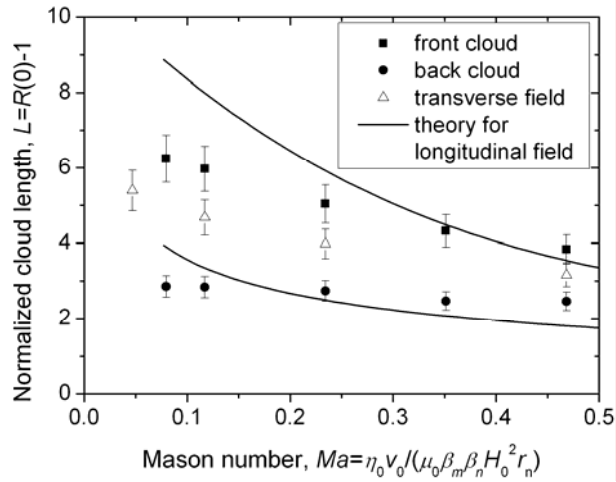

Fig.9. Theoretical and experimental dependencies of the normalized cloud length on nanocluster Mason number. The upper solid line stands for the theoretical prediction (without adjustable parameters) for the front cloud in the longitudinal field [cf. Eqs. (7), (11a)]. The lower solid line stands for the prediction for the back cloud in the longitudinal field; the best fit to the experimental data for the back cloud corresponds to the adjustable parameter $c \approx 1.6$ [cf. Eqs. (7), (11b)]

Finally, our theory reveals that the magnetic interactions between nanoclusters dominate over the Brownian motion at the field parameters $\alpha \geq 2$ and the nanocluster magnetic permeability $\mu_n \geq 30$. At this range of both parameters, the Brownian motion seems to not alter significantly the size, the shape and the internal volume fraction of the clouds. The single dimensionless parameter affecting the clouds is therefore the Mason number $Ma = \eta_0 v_0 /(\mu_0 \beta_m \beta_n H_0^2 r_n)$. The ratio of the hydrodynamic-to-magnetic stresses, $\sigma_h / \sigma_m = \eta_0 v_0 /(\mu_0 H_0^2 r_m)$, might govern the cloud shape. However, it appears to be smaller by a factor of $r_m / r_n \sim 10^3$ than the Mason number (at $\beta_m \approx \beta_n \approx 1$) and does not seem to affect the cloud behavior at the considered range of the flow speeds. These conclusions hold true for nanoclusters or nanoparticles with $\mu_n \geq 30$. A separate study should be carried out to check the role of the parameter $\mu_n$ on the phase transition and cloud behavior under flow.

## V. Conclusions

In the present work, size and shape of the clouds of magnetic nanoparticles (or nanoclusters) around a magnetized microparticle in the presence of an external uniform magnetic field and in the presence of flow have been studied in details. In experiments, we have used nanoclusters composed of numerous iron oxide nanoparticles and having a mean size of about 60 nm and the initial magnetic permeability $\mu_n \approx 30$. These nanoclusters were accumulated around a spherical nickel microparticle of a mean diameter of 50 μm. The main results of this study can be summarized as follows:

1. Even in the absence of microparticles, the external magnetic field induces rather strong interactions between magnetic nanoclusters, such that the whole nanocluster suspension may undergo a phase separation, which is manifested by appearance of needle-like aggregates composed of nanoclusters. In our experimental case, this phase separation has been observed in the suspension bulk at volume fraction of solids as low as $\phi=0.32\%$ (corresponding to the concentration of nanoclusters, $\Phi=\phi/\Phi_n \approx 0.64\%$) and at a magnetic field $H_0=12$ kA/m. Such a phase separation is expected to be a signature of a "disordered fluid – ordered solid" phase transition. Assuming FCC-lattice of the solid phase and multipolar magnetic interactions between the nanoclusters, we have constructed a phase diagram of the magnetic suspension and shown that the considered phase transition is governed by the dipolar coupling parameter $\alpha = \mu_0 \beta_n^2 H_0^2 V_n /(2kT)$, the nanocluster concentration, $\Phi$, and the nanocluster magnetic permeability $\mu_n$. The theory confirms the phase separation at the considered experimental parameters ($\alpha > 2$, $\Phi \approx 0.64\%$ and $\mu_n \approx 30$).

2. The above stated phase separation appears to be extremely important for the efficiency of capture of nanoclusters by a magnetized microparticle. In the absence of flow and in the presence of an external magnetic field, the nanoclusters are attracted to the microparticle and form two equal-sized clouds attached to both magnetic poles of the microparticle and aligned with the applied field. Phase equilibrium analysis reveals a solid state of the matter in the clouds with nanocluster volume fraction close to the maximum packing fraction of the FCC lattice, $\Phi_m \approx 0.74$. To achieve the equilibrium, the clouds are expected to absorb the major amount of nanoclusters (and their aggregates) from the suspension bulk that would result in extremely large cloud size. In practice, the cloud growth is limited by inevitable settling of the nanocluster aggregates, which adhere to the wall of the experimental cell and cannot move.

3. In the presence of strong enough flow, the settling problem may be overcome so that the nanoclusters (and their aggregates) continuously arrive to the microparticle. Starting from some speed, at which the travel time becomes smaller than the settling time, the size of the clouds monotonically decreases with increasing flow speed in both longitudinal and transverse fields. This is qualitatively explained by enhancement of hydrodynamic forces washing the nanoclusters away from the clouds. In the longitudinal field, the flow induces asymmetry of the front and the back clouds, which likely comes from stronger hydrodynamic forces acting on the nanoclusters on the surface of the back cloud (near its rear point) as compared to those acting on the front cloud (near the anchoring point).

4. In the transverse field, both clouds are of equal size but seem to be slightly misaligned with the field direction in the direction of flow. Conical spikes appear on the cloud extremities at both field orientations. This phenomenon is interpreted in terms of the surface energy anisotropy, which induces surface instabilities in magnetic fields making the angles $\delta$>31 deg with the surface. Finally, the cloud formation / destruction is completely reversible process: when the magnetic field is switched off, the clouds are rapidly destroyed by the suspension flux and by Brownian motion of nanoclusters.

5. To explain the flow and the field effects on the clouds, we have developed a simple model ignoring appearance of spikes and based on the balance of the stresses and particle fluxes on the cloud surface. This model, applied to the case of the magnetic field parallel to the flow, captures reasonably well the elongated shape of the cloud and reveals that the only dimensionless parameter governing the cloud size is the Mason number, $Ma = \eta_0 v_0 / (\mu_0 \beta_m \beta_n H_0^2 r_n)$. At strong magnetic interactions considered in the present work ($\alpha \geq 2$ and $\mu_n \geq 30$), the Brownian motion seems not to affect the cloud behavior. The model correctly predicts a decreasing Mason number dependency of the cloud size and allows obtaining a satisfactory (within maximum 35% error) agreement with experiments without adjustable parameters for the front cloud and with a single fitting parameter for the back cloud.

In summary, the present theory captures the main physics of the cloud behavior and provides tractable semi-analytical results at relatively low computational expense. For the better agreement between theory and experiments, more precise calculations of magnetic field, concentration and velocity distribution are needed. New experiments should be carried out to analyze the effect of the initial nanocluster concentration on the cloud size and shape. Unsteady-state of the nanocluster accumulation should also be studied in order to estimate characteristic times of the cloud formation. From an application perspective, we intend to realize a micro-separation device composed of a microfluidic flow channel containing a regular array of magnetizable microplots. Such a channel has already been used by Deng et al. [53] as on-chip cell sorting device with the cells bound to the magnetic micron-sized beads captured by magnetized microplots. We intend to study the separation of smaller nano-sized particles in such a device and visualize the nanoparticle fluxes by a fluorescence microscopy. The interference between the flow fields as well as between magnetic fields generated around neighboring microplots is expected to considerably change the physics of the nanoparticle capture.

**Acknowledgements**

This work has been supported by the project "Factories of the Future" (grant No. 260073, DynExpert FP7), by the Russian Fund of Fundamental Investigations (grant No. 13-02-91052), by the project PICS 6102 CNRS/Ural Federal University and by the Belarusian Republican Foundation for Fundamental Research (exchange of scientists grant 2013-2014).

**Appendix. Estimation of the anchoring angle [Eqs. (11a), (11b)]**

The anchoring angle of both clouds is found from the force balance (10) applied to the characteristic points A and B on the cloud surface [cf. Fig. 6]. We proceed to these estimations separately for the point A on the front cloud and the point B on the back cloud.

*(a) Front cloud.*

The normal component of the magnetic force $(F_m)_n$, intervening into Eq. (10), is related to the normal component, $\nabla_n H$, of the magnetic field gradient by the following expression:

$$(F_m)_n = \mu_0 m \nabla_n H = \frac{3}{2}\mu_0 \beta_n V_n \nabla_n H^2 \tag{A1}$$

The magnetic field and its gradient at the point A are estimated using the well-known expressions for the magnetic field distribution around an isolated sphere placed into a non-magnetic medium (c.f. expressions for $h_r$ and $h_\theta$ below Eq. 7). This gives the following estimate of the force $(F_m)_n$:

$$(F_m)_n = -12\pi\mu_0 \beta_m \beta_n (2+\beta_m)\frac{H_0^2}{r_m} r_n^3 \cos\theta_0 \sin\theta_0 \tag{A2}$$

The normal component $\left(F_{m,eq}\right)_n$ of the magnetic force in equilibrium, i.e. in the absence of flow, is calculated at the anchoring point A for the same position of the cloud surface as in the absence of flow. This position corresponds to the anchoring angle $\theta_0 = \theta_{eq} \approx 69$ deg, as explained below Eqs. (11). The expression for $\left(F_{m,eq}\right)_n$ reads:

$$\left(F_{m,eq}\right)_n = -12\pi\mu_0 \beta_m \beta_n (2+\beta_m)\frac{H_0^2}{r_m} r_n^3 \cos\theta_{eq} \sin\theta_{eq} \tag{A3}$$

At the anchoring point A, the normal component $v_n$ of the fluid velocity at the cloud surface appears to be the tangential component $v_\theta$ at the microparticle surface. It is estimated at a distance equal to the nanocluster radius $r_n$ from the cloud surface, so as a product of the shear rate $\dot\gamma$ at the microparticle surface and the nanocluster radius. The shear rate is roughly estimated using the well-known result for the velocity distribution around an isolated sphere in the absence of cloud [54]. The final expression for the velocity $v_n$ reads:

$$v_n = v_\theta \sim \dot\gamma r_n \sim \frac{3}{2}\frac{v_0}{r_m} r_n \sin\theta_0 \tag{A4}$$

Replacing Eqs. (A2)-(A4) in the force balance (10), we arrive, after some algebra, to the equation (11a) describing the anchoring angle of the front cloud.

*(b) Back cloud*

The point B is situated at the extremity of the back cloud. As explained in Sec. III, the negative surface energy forbids existence of smooth surfaces perpendicular to the magnetic field, therefore the cloud extremity B is a singular point of the cloud surface. The radius of curvature of the cloud surface at this point is expected to be of the order of magnitude of a few nanocluster radii, $r_n$. This implies strong magnetic field gradients at this point. At the present time we are only able to estimate the order of magnitude of the field gradient at this point:

$$\nabla_n H^2 \sim \frac{\beta_m H_B^2}{r_n} \quad (A5)$$

where $H_B$ is the magnetic field intensity at the point B. An order of magnitude of the field $H_B$ is given by its value at the point B in the absence of cloud:

$$H_B \sim H_0 \left(1 + \frac{2\beta_m}{R_B^3(\theta_0)}\right) \quad (A6)$$

with $R_B(\theta_0)$ being a distance between the microsphere center and the point B on the extremity of the back cloud. Combining Eqs. (A1), (A5) and (A6), we obtain an estimation of the magnetic force acing on nanoclusters at the point B under flow, $(F_m)_n$, and in equilibrium, $(F_{m,eq})_n$:

$$(F_m)_n \sim \mu_0 \beta_m \beta_n H_0^2 \left(1 + \frac{2\beta_m}{R_B^3(\theta_0)}\right)^2 r_n^2 \quad (A7)$$

$$(F_{m,eq})_n \sim \mu_0 \beta_m \beta_n H_0^2 \left(1 + \frac{2\beta_m}{R_B^3(\theta_{eq})}\right)^2 r_n^2 \quad (A8)$$

The magnetic forces under flow [Eq. (A7)] and in the absence of flow [Eq. (A8)] are calculated using the appropriate values $\theta_0(Ma)$ and $\theta_{eq}$ of the anchoring angle. The strain rate $|\nabla \mathbf{v}|$ of the fluid in the vicinity of the point B is governed by the radius of curvature of the surface at this point. This allows estimation of the order of magnitude of the speed $v_n$ near the extremity of the back cloud, at a distance $r_n$ from the point B:

$$v_n \approx |\nabla \mathbf{v}| r_n \sim \left(\frac{v_0}{r_n}\right) r_n \sim v_0 \quad (A9)$$

Replacing Eqs. (A7)-(A9) in Eq. (10), we arrive, after some algebra, to the equation (11b) describing the anchoring angle of the back cloud.